# Optimum multiplexer design in quantum-dot cellular automata

Esam AlKaldy[1], Ali H. Majeed[2], Mohd Shamian bin Zainal[3], Danial Bin MD Nor[4]
[1,2]Electrical Engineering, Collage of Engineering, University of Kufa, Iraq
[2,3,4]Electrical and Electronic Engineering, UTHM, Malaysia

| Article Info | ABSTRACT |
|---|---|
| *Article history:*<br><br>Received Apr 13, 2019<br>Revised Jun 25, 2019<br>Accepted Jul 9, 2019<br><br>*Keywords:*<br><br>Multiplexer<br>Nanotechnology<br>QCA<br>Quantum-dot cellular automata | Quantum-dot Cellular Automata (QCA) is one of the most important computing technologies for the future and will be the alternative candidate for current CMOS technology. QCA is attracting a lot of researchers due to many features such as high speed, small size, and low power consumption. QCA has two main building blocks (majority gate and inverter) used for design any Boolean function. QCA also has an inherent capability that used to design many important gates such as XOR and Multiplexer in optimal form without following any Boolean function. This paper presents a novel design 2:1 QCA-Multiplexer in two forms. The proposed design is very simple, highly efficient and can be used to produce many logical functions. The proposed design output comes from the inherent capabilities of quantum technology. New 4:1 QCA-Multiplexer has been built using the proposed structure. The output waveforms showed the wonderful performance of the proposed design in terms of the number of cells, area, and latency.<br><br> |

*Corresponding Author:*

Ali H. Majeed,
Electrical Engineering,
Collage of Engineering,
University of Kufa, 54001, Kufa, Iraq.
Email: Alih.alasady@uokufa.edu.iq

## 1. INTRODUCTION

In recent years, there have been many studies to replace CMOS technology that currently used in Integrated Circuits (ICs) with an alternative because of many reasons such as the power consumption and inability to continue following Moore's law [1], which states the number of transistors in a single chip. So, searching for alternatives became a priority for the researchers. QCA technology has many amazing features such as high speed, low complexity, and small size compared with traditional CMOS. Quantum cell is the basic building block in QCA circuits, each cell has four quantum dots injected with two electrons moving between those dots [2]. QCA uses the arrangement of intracellular electrons for binary representation instead of voltage level, so there is flow of current in this technique [3-5]. Many efforts have been made to implement the crucial components in logic circuits such as XOR [6-11] and multiplexer [12-18] using this nanotechnology. The multiplexer (MUX) was extensively utilized to implement many digital circuits such as RAM cells and ALU. The reliability of QCA circuit got the attention in [19]. This paper presents a novel structure 2:1 QCA-MUX. The proposed design has many enhancements in terms of area, latency, and complexity. The 4:1 multiplexer is carried out using the proposed structure.

Quantum cell is the brick in QCA circuits. Every cell, Square-formed, have four quantum dots. Two electrons are injected into each cell and these electrons have the ability to move between points so that they settle in a diagonal position because of the columbic repulsion dependence on driver cell [20]. Figure 1 illustrates cell polarization. Binary numbers 1 and 0 can be represented with the two polarizations of cell P = +1 and P= –1 respectively. QCA wire and logical functions can be implemented by forming a group of cells in an array.

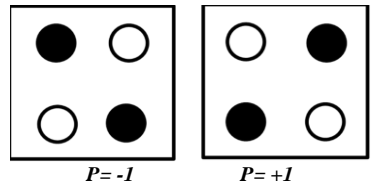

Figure 1. Cell Polarization

QCA wire is used to carry out the binary data transformation from the input cell to the output. Since the wire in the QCA consists of an array of cells, the binary data will be transferred to the output cell according to the principle of electronic repulsion [21]. The QCA wire is presented in two configurations; normal and rotated, as explained in Figure 2. To achieve coplanar wire crossing, the rotated wire is required or use another approach which introduced in [22] since there are other technique uses multi-layer but it is not promises for physical implementation.

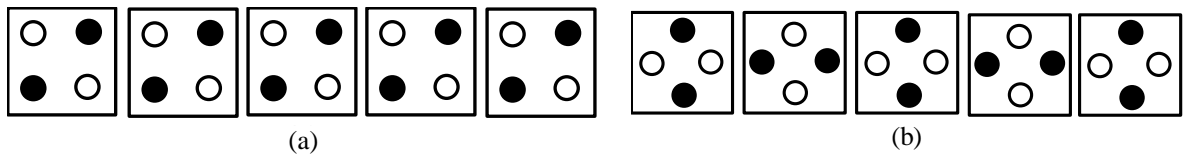

Figure 2. QCA wire (a) normal wire, (b) rotated wire

The main logic gates AND and OR can be performed utilizing the QCA universal gate (majority gate) by setting one of the inputs to 0 and 1 respectively. Majority gate is dominant in the QCA world with several studies focusing on it such as [23-27]. Two configurations of the majority gate are introduced in QCA as illustrated in Figure 3 [28]. The formula of the majority gate is given by 1.

$$M(A,B,C)=AB+BC+AC \tag{1}$$

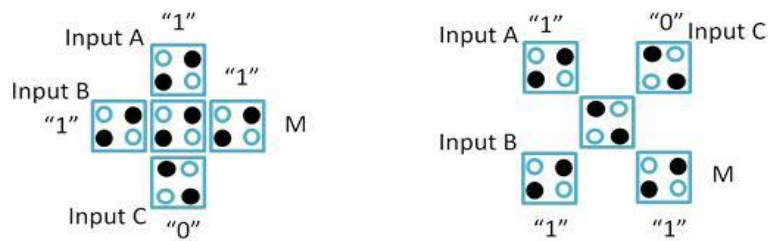

Figure 3. Majority gate forms

Generally, inverter with majority gate represents the fundamental blocks in QCA circuits, three forms of inverter were introduced in QCA as shown in Figure 4.

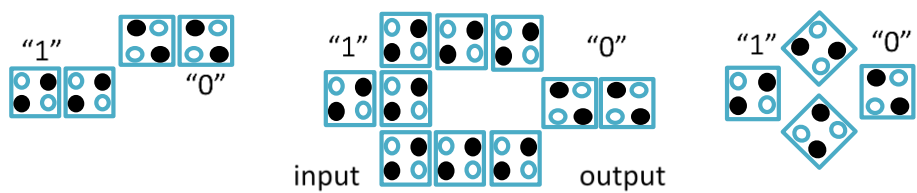

Figure 4. QCA inverter forms

To ensure data flow from input to output and to ensure cells synchronization, the clock is applied to all cells [20]. The barriers between the dots inside the cells are controlled by the clocking signal so that the cell is unpolarized as long as the clock is low. When the clock goes high, each cell gets polarized after the electrons move to the dots which need the lowest energy depending on the driver cell. Clock signal consists of four phases to guarantee adiabatic cell switching, (switch, hold, release and relax). QCA circuit can be divided into 4 zones where every zone comprises four phases as illustrated in Figure 5 [29].

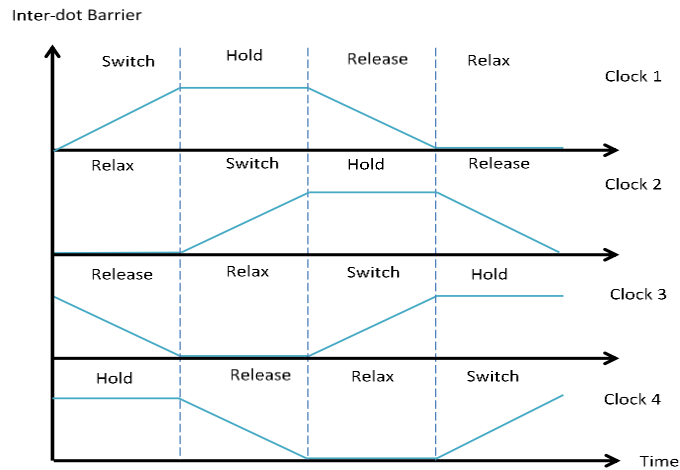

Figure 5. QCA Clock signal

Multiplexer circuit takes information from 2n lines where at each time one of the inputs send to the output depending on selector signals [13]. Table 1 shows the 2:1 multiplexer truth table. If S is 0 then the value of I0 goes to the output and if S is 1 then the value of I1 goes to the output. The output equation of multiplexer is given in 2. Figure 6. shows the previously proposed designs of 2:1 MUX. The first three structures consist of 2-input AND gates, 2-input OR gate and inverter as the schematic diagram is shown in Figure 7. [3] while the last structure uses the inherent capabilities of QCA to get the output. The design in Figure 6 (a) uses 0.03 μm2 and consists of 27 cells with 3 clock zone latency. The design in Figure 6 (b) uses 0.02 μm2 and consists of 26 cells with 2 clock zone latency. The design in Figure 6 (c) uses 0.02 μm2 and consists of 23 cells with 2 clock zone latency. While the design in Figure 6 (d) uses 0.01 μm2 and consists of 12 cells with 1 clock zone latency.

$$Out = S.I_1 + \overline{S}.I_0 \qquad (2)$$

Table 1. 2:1 Mux Truth Table

| S | Out |
|---|-----|
| 0 | $I_0$ |
| 1 | $I_1$ |

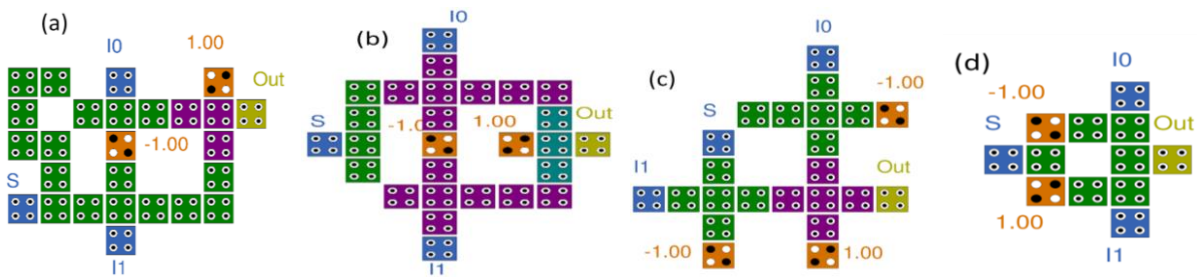

Figure 6. 2:1 MUX presented (a) in [30], (b) in[31], (c) in[32]and (d) in [13]

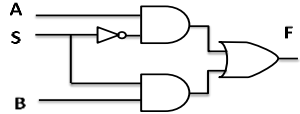

Figure 7. Logic circuit diagram of 2:1 Multiplexer

## 1.1. The Proposed Circuits
This section divided into two parts, the first part presents tow new structures of 2:1 Multiplexer, that can be interchanged depending on the circuit requirements, and in the second part new 4:1 multiplexer is implemented using the unique structure which was proposed in the first part.

## 1.2. Proposed 2:1 Multiplexer
The schematic of the multiplexer and the layouts of the proposed two circuits are shown in Figure 8. (a), (b) and (c) respectively. It contains two inputs (I0, I1), one output (Out) and one selector (S).

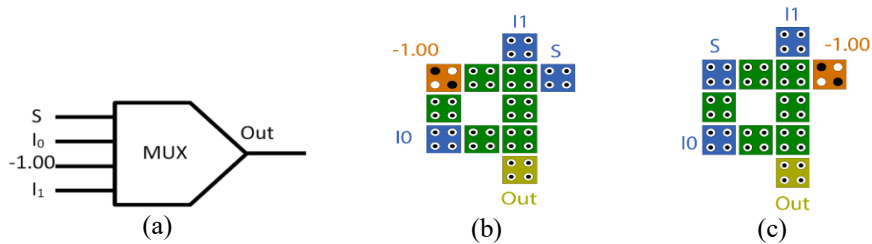

Figure 1. Proposed 2:1 QCA-MUX (a) Schematic, (b) Layout 1, (c) Layout 2

## 1.3. Proposed 4:1 Multiplexer
The 4:1 MUX contains 4 inputs ($I_0$, $I_1$, $I_2$, and $I_3$), 1 output (Out) and 2 selectors ($S_0S_1$). The truth table of this Multiplexer is illustrated in Table 2. The Out equal to $I_0$ when $S_0S_1$=00, when $S_0S_1$=01 the Out =$I_1$ however, the Out equal to $I_2$ and $I_3$ if $S_0S_1$ equal to 10 and 11 respectively. The 4:1 Multiplexer constructed of three 2:1 multiplexer and the formula of this multiplexer is given in 3. Figure 9 shows the schematic and structure layout of the proposed 4:1 multiplexer.

Table 2. 4:1 QCA-Mux Truth Table

| $S_0S_1$ | Out |
|---|---|
| 00 | $I_0$ |
| 01 | $I_1$ |
| 10 | $I_2$ |
| 11 | $I_3$ |

$$\text{Out} = (\overline{S_0}\,\overline{S_1})I_0 + (\overline{S_0}S_1)I1 + (S_0\overline{S_1})I2 + (S_0S_1)I3 \tag{3}$$

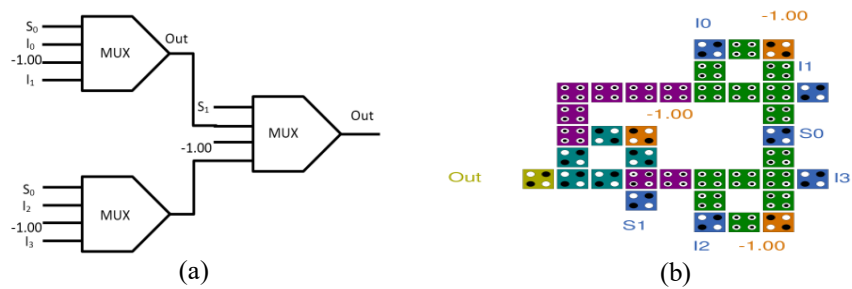

Figure 9. Proposed 4:1 MUX (a) Schematic, (b) Layout

## 2. METHODOLOGY

The Proposed multiplexers are verified by QCADesigner V 2.0.3, with the default simulation parameters. Figure 10 shows the simulation result of the proposed 2:1 QCA-MUX. The output waveforms proved that they were identical to what is expected without delay because input and output are in the same clock zone. The proposed unique 2:1 multiplexer uses 0.01 μm$^2$ and consists of 11 cells with 1 clock zone latency. The result of the simulation for the second proposed structure, 4:1 multiplexer, is illustrated in Figure 11. This QCA circuit consists of 37 cells with 0.03 μm$^2$ and 3 clock zone latency. This result is extremely distinguished in QCA multiplexer designs.

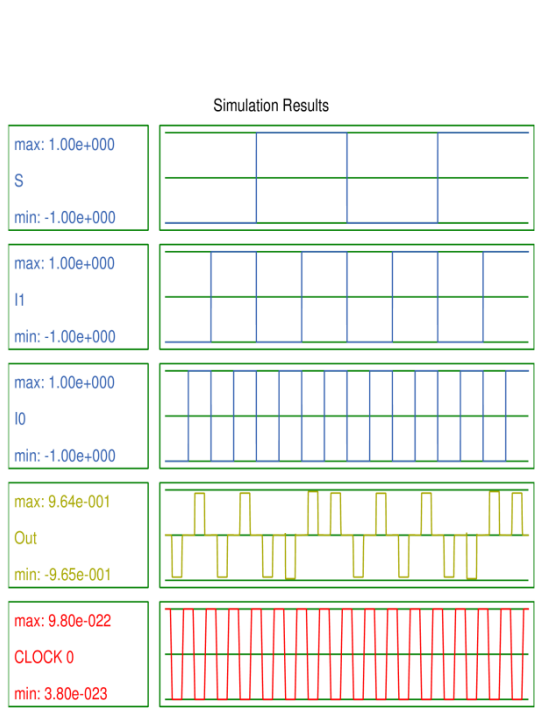

Figure 10. Output waveforms for the proposed 2:1 MUX

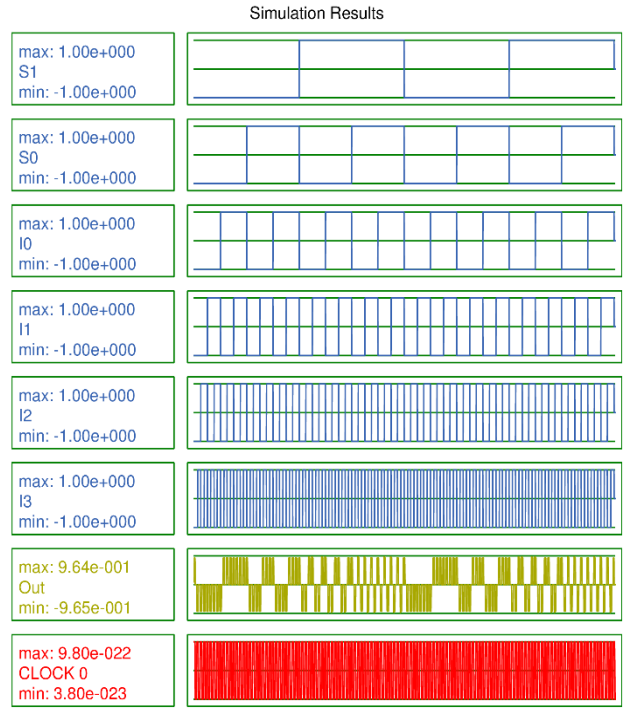

Figure 11. Output waveforms for the proposed 4:1 MUX

## 3. RESULTS AND DISCUSSION

This section shows the performance of the proposed circuits by comparing the result with previous designs. Table 3 illustrate the comparison result for the first MUX, 2:1 multiplexer, proposed with existing counterparts. While the comparison results of the second proposed gate, 4:1 multiplexer, is shown in Table 4.

Table 3. Comparative Result of 2:1 Multiplexer

| 2:1 Mux | Area (μm$^2$) | No. of Cells | Crossover type | Latency (Clock zone) |
|---|---|---|---|---|
| [33]Fig. 9 | 0.28 | 146 | Multilayer | 8 |
| [33]Fig. 8 | 0.14 | 88 | Multilayer | 4 |
| [34] | 0.14 | 67 | Coplanar | 4 |
| [35] | 0.07 | 56 | Coplanar | 4 |
| [36] | 0.08 | 46 | Coplanar | 4 |
| [15] | 0.06 | 36 | Multilayer | 4 |
| [37] | 0.04 | 35 | Coplanar | 4 |
| [38] | 0.03 | 27 | Coplanar | 3 |
| [31] | 0.02 | 26 | Coplanar | 2 |
| [32] | 0.02 | 23 | Coplanar | 2 |
| [16] | 0.02 | 19 | Coplanar | 2 |
| [39] | 0.02 | 19 | Coplanar | 3 |
| [3] | 0.01 | 15 | Coplanar | 2 |
| [13] | 0.01 | 12 | Without | 1 |
| proposed | 0.01 | 11 | Without | 1 |

Table 4. Comparative Result of 4:1 Multiplexer

| 4:1 Mux | Area (μm²) | No. of Cells | Crossover type | Latency (Clock zone) |
|---|---|---|---|---|
| [35] | 0.35 | 290 | Coplanar | 6 |
| [31] | 0.37 | 271 | Coplanar | 19 |
| [40] | 0.2 | 251 | Multilayer | 5 |
| [41] | 0.22 | 223 | Multilayer | 6 |
| [34] | 0.25 | 215 | Coplanar | 6 |
| [40] | 0.27 | 199 | Coplanar | 6 |
| [32] | 0.24 | 155 | Coplanar | 5 |
| [37] | 0.25 | 124 | Coplanar | 8 |
| [3] | 0.15 | 107 | Coplanar | 4 |
| [13] | 0.08 | 61 | Coplanar | 4 |
| Proposed | 0.03 | 37 | Without | 3 |

The power dissipation of the above circuits including the two layouts of the proposed multiplexer can be estimated using QCAPro tool. This tool capable of dealing with a large number of cells because it utilizes a fast approximation-based technique and it can expect non-adiabatic switching power losses with polarization error in QCA circuit. In this work, the value of temperature 2k has been taken in QCAPro parameter. The comparative analysis of dissipated power at different levels of tunneling energy (0.5Ek, 1Ek, and 1.5Ek) for 2:1 Multiplexer are shown in Table 5. The maps of dissipated power for the presented circuits with (0.5Ek) tunneling energy are illustrated in Figure 12.

Table 5. Power Analysis Result

| Circuit | Avg. leakage energy dissipation (meV) | | | Avg. switching energy dissipation (meV) | | | Total energy consumption (meV) | | |
|---|---|---|---|---|---|---|---|---|---|
| | 0.5Ek | 1Ek | 1.5Ek | 0.5Ek | 1Ek | 1.5Ek | 0.5Ek | 1Ek | 1.5Ek |
| In [30] | 7.36 | 21.78 | 38.79 | 32.40 | 28.23 | 24.20 | 39.76 | 50.01 | 62.99 |
| In [31] | 8.19 | 23.80 | 41.82 | 29.15 | 25.06 | 21.21 | 37.34 | 48.86 | 63.03 |
| In [32] | 7.23 | 20.54 | 35.68 | 24.37 | 20.77 | 17.48 | 31.60 | 41.31 | 53.16 |
| In [13] | 3.43 | 9.61 | 16.46 | 11.54 | 9.50 | 7.86 | 14.97 | 19.11 | 24.32 |
| Proposed1 | 2.66 | 7.72 | 13.59 | 11.15 | 9.90 | 8.63 | 13.81 | 17.62 | 22.22 |
| Proposed2 | 2.54 | 7.50 | 13.35 | 8.07 | 7.04 | 6.07 | 10.61 | 14.54 | 19.42 |

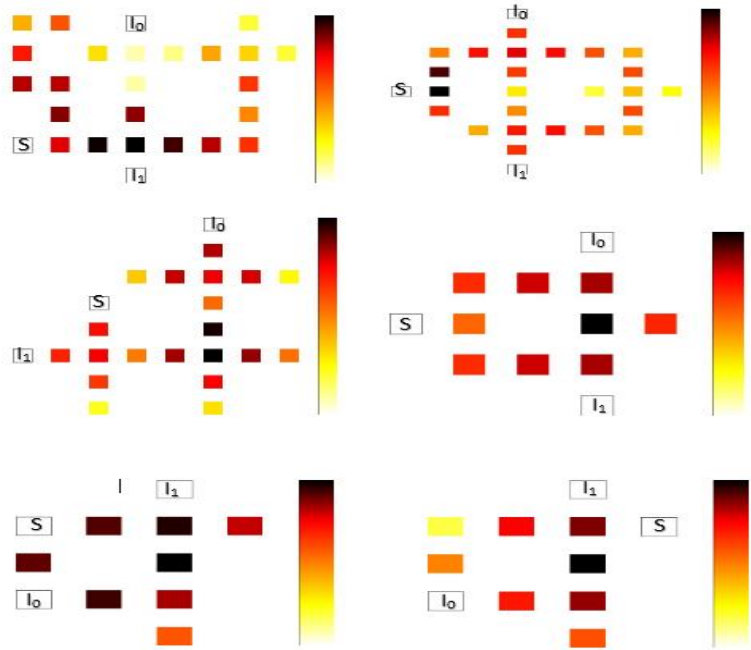

Figure 12. Dissipated power maps for multiplexer presented (a) in [30], (b) in[31], (c) in[32], (d) in [13], (e) proposed layout 1, (f) proposed layout 2

*Optimum multiplexer design in quantum-dot cellular automata (Esam AlKaldy)*

## 4. CONCLUSION

This paper proposes a novel highly efficient two forms singular 2:1 QCA-MUX circuit simulated by QCADesigner tool. The proposed gate does not follow any logical functions. It is derived from the inherent capabilities of quantum technology to give the correct output. A new structure of 4:1 multiplexer has been designed with coherence vector engine using the proposed 2:1 multiplexer. The unique feature for the 4:1 multiplexer is that it was done without any crossover. The proposed designs prove that it has less complexity as well as being cost effective compared to the typical multiplexers. The simulation results obtained from the QCADesigner tool show that the structures presented in this work show an improvement in terms of area, delay and cell count putting in mind that all inputs and outputs are in the circuit terminals which gives real single layer design without wire crossing.

**REFERENCES**
[1] G. E. Moore, "Cramming more components onto integrated circuits, Reprinted from Electronics, volume 38, number 8, April 19, 1965, pp.114 ff," *IEEE Solid-State Circuits Society Newsletter,* vol. 11, pp. 33-35, 2006.
[2] C. S. Lent et al, "Quantum_cellular_automata," *Nanotechnology,* vol. 4, pp. 49-57, 1993.
[3] H. Rashidi, A. Rezai, and S. Soltany, "High-performance multiplexer architecture for quantum-dot cellular automata," *Journal of Computational Electronics,* vol. 15, pp. 968-981, 2016/09/01 2016.
[4] V. Nath, P. K. Barhai, and D. K. Verma, "QCA and CMOS Nanotechnology Based Design and Development of Nanoelectronic Security Devices with Encryption Schemes," *TELKOMNIKA Indonesian Journal of Electrical Engineering,* pp. 270-279, 2015.
[5] W. Musa, S. Wahyuni Dali, and A. Irawaty Tolago, *Design of Digital Parity Generator Layout Using 0.7 micron Technology* vol. 8, 2018.
[6] P. Z. Ahmad, F. Ahmad, and H. A. Khan, "A new F-shaped XOR gate and its implementations as novel adder circuits based Quantum-dot cellular Automata (QCA)," *IOSR Journal of Computer Engineering (IOSR-JCE),* vol. 16, p. 2014, 2014.
[7] M. R. Beigh, M. Mustafa, and F. Ahmad, "Performance Evaluation of Efficient XOR Structures in Quantum-Dot Cellular Automata (QCA)," *Circuits and Systems,* vol. 04, pp. 147-156, 2013.
[8] D. AJITHA, K. V. RAMANAIAH, and V. SUMALATHA, "AN EFFICIENT DESIGN OF XOR GATE AND ITS APPLICATIONS USING QCA," *i-manager's Journal on Electronics Engineering,* vol. 5, pp. 22-29, 2015.
[9] G. Singh, R. K. Sarin, and B. Raj, "A novel robust exclusive-OR function implementation in QCA nanotechnology with energy dissipation analysis," *Journal of Computational Electronics,* vol. 15, pp. 455-465, 2016/06/01 2016.
[10] A. N. Bahar, S. Waheed, N. Hossain, and M. Asaduzzaman, "A novel 3-input XOR function implementation in quantum dot-cellular automata with energy dissipation analysis," *Alexandria Engineering Journal,* 2017.
[11] M. Poorhosseini and A. R. Hejazi, "A Fault-Tolerant and Efficient XOR Structure for Modular Design of Complex QCA Circuits," *Journal of Circuits, Systems and Computers,* vol. 27, p. 1850115, 2018.
[12] A. M. Chabi, S. Sayedsalehi, S. Angizi, and K. Navi, "Efficient QCA Exclusive-or and Multiplexer Circuits Based on a Nanoelectronic-Compatible Designing Approach," *Int Sch Res Notices,* vol. 2014, p. 463967, 2014.
[13] M. Naji Asfestani and S. Rasouli Heikalabad, "A unique structure for the multiplexer in quantum-dot cellular automata to create a revolution in design of nanostructures," *Physica B: Condensed Matter,* vol. 512, pp. 91-99, 2017.
[14] F. Ahmad, "An optimal design of QCA based 2 n :1/1:2 n multiplexer/demultiplexer and its efficient digital logic realization," *Microprocessors and Microsystems,* vol. 56, pp. 64-75, 2018.
[15] S. Hashemi, M. R. Azghadi, and A. Zakerolhosseini, "A novel QCA multiplexer design," in *2008 International Symposium on Telecommunications*, 2008, pp. 692-696.
[16] B. Sen, M. Dutta, M. Goswami, and B. K. Sikdar, "Modular Design of testable reversible ALU by QCA multiplexer with increase in programmability," *Microelectronics Journal,* vol. 45, pp. 1522-1532, 2014/11/01 2014.
[17] M. Goswami, B. Kumar, H. Tibrewal, and S. Mazumdar, "Efficient realization of digital logic circuit using QCA multiplexer," in *2014 2nd International Conference on Business and Information Management (ICBIM)*, 2014, pp. 165-170.
[18] B. Sen, A. Nag, A. De, and B. K. Sikdar, "Multilayer design of QCA multiplexer," in *2013 Annual IEEE India Conference (INDICON)*, 2013, pp. 1-6.
[19] V. Vankamamidi, M. Ottavi, and F. Lombardi, "A line-based parallel memory for QCA implementation," *IEEE Transactions on Nanotechnology,* vol. 4, pp. 690-698, 2005.
[20] K.-m. Qiu and Y.-s. Xia, "Quantum-dots cellular automata comparator," in *2007 7th International Conference on ASIC*, 2007, pp. 1297-1300.
[21] M. B. Khosroshahy, M. H. Moaiyeri, K. Navi, and N. Bagherzadeh, "An energy and cost efficient majority-based RAM cell in quantum-dot cellular automata," *Results in Physics,* vol. 7, pp. 3543-3551, 2017.
[22] S. Angizi, E. Alkaldy, N. Bagherzadeh, and K. Navi, "Novel Robust Single Layer Wire Crossing Approach for Exclusive OR Sum of Products Logic Design with Quantum-Dot Cellular Automata," *Journal of Low Power Electronics,* vol. 10, pp. 259-271, 2014.
[23] M. A. Tehrani, K. Navi, and A. Kia-kojoori, "Multi-output majority gate-based design optimization by using evolutionary algorithm," *Swarm and Evolutionary Computation,* vol. 10, pp. 25-30, 2013.

[24] R. Zhang, P. Gupta, and N. K. Jha, "Majority and Minority Network Synthesis With Application to QCA-, SET-, and TPL-Based Nanotechnologies," *IEEE Transactions on Computer-Aided Design of Integrated Circuits and Systems,* vol. 26, pp. 1233-1245, 2007.
[25] M. R. Bonyadi, S. M. R. Azghadi, N. M. Rad, K. Navi, and E. Afjei, "Logic Optimization for Majority Gate-Based Nanoelectronic Circuits Based on Genetic Algorithm," in *2007 International Conference on Electrical Engineering*, 2007, pp. 1-5.
[26] Ali H. Majeed, E. AlKaldy, MSB Zainal, and Danial BMD Nor, "A new 5-input Majority Gate Without Adjacent Inputs Crosstalk Effect in QCA Technology," *Indonesian Journal of Electrical Engineering and Computer Science,* vol. 14, 2019.
[27] E. Alkaldy, K. Navi, F. Sharifi, and M. Moaiyeri, *An Ultra High-Speed (4;2) Compressor with a New Design Approach for Nanotechnology Based on the Multi-Input Majority Function* vol. 11, 2014.
[28] M. Bagherian Khosroshahy, M. Hossein Moaiyeri, and K. Navi, *Design and evaluation of a 5-input majority gate-based content-addressable memory cell in quantum-dot cellular automata*, 2017.
[29] F. Ahmad, G. M. Bhat, and P. Z. Ahmad, "Novel Adder Circuits Based On Quantum-Dot Cellular Automata (QCA)," *Circuits and Systems,* vol. 05, pp. 142-152, 2014.
[30] A. Roohi, H. Khademolhosseini, S. Sayedsalehi, and K. Navi, "A Novel Architecture for Quantum-Dot Cellular Automata Multiplexer," 2011.
[31] R. Sabbaghi-Nadooshan and M. Kianpour, "A novel QCA implementation of MUX-based universal shift register," *Journal of Computational Electronics,* vol. 13, pp. 198-210, 2014/03/01 2014.
[32] B. Sen, M. Goswami, S. Mazumdar, and B. K. Sikdar, "Towards modular design of reliable quantum-dot cellular automata logic circuit using multiplexers," *Computers & Electrical Engineering,* vol. 45, pp. 42-54, 2015.
[33] T. Teodosio and L. Sousa, "QCA-LG: A tool for the automatic layout generation of QCA combinational circuits," in *Norchip 2007*, 2007, pp. 1-5.
[34] V. Mardiris, C. Mizas, L. Fragidis, and V. Chatzis, "Design and simulation of a QCA 2 to 1 multiplexer," presented at the Proceedings of the 12th WSEAS international conference on Computers, Heraklion, Greece, 2008.
[35] V. A. Mardiris and I. Karafyllidis, "Design and simulation of modular 2n to 1 quantum-dot cellular automata (QCA) multiplexers," *I. J. Circuit Theory and Applications,* vol. 38, pp. 771-785, 2010.
[36] K. Kim, K. Wu, and R. Karri, "The Robust QCA Adder Designs Using Composable QCA Building Blocks," *IEEE Transactions on Computer-Aided Design of Integrated Circuits and Systems,* vol. 26, pp. 176-183, 2007.
[37] M. Askari, M. Taghizadeh, and K. Fardad, "Digital design using quantum-dot cellular automata (A nanotechnology method)," in *2008 International Conference on Computer and Communication Engineering*, 2008, pp. 952-955.
[38] K. Walus, T. J. Dysart, G. A. Jullien, and R. A. Budiman, "QCADesigner: a rapid design and Simulation tool for quantum-dot cellular automata," *IEEE Transactions on Nanotechnology,* vol. 3, pp. 26-31, 2004.
[39] B. Sen, M. Dutta, D. Saran, and B. K. Sikdar, "An Efficient Multiplexer in Quantum-dot Cellular Automata," in *Progress in VLSI Design and Test*, Berlin, Heidelberg, 2012, pp. 350-351.
[40] G. Cocorullo, P. Corsonello, F. Frustaci, and S. Perri, "Design of efficient QCA multiplexers," *International Journal of Circuit Theory and Applications,* vol. 44, pp. 602-615, 2016/03/01 2016.
[41] V. Vankamamidi, M. Ottavi, and F. Lombardi, "Two-Dimensional Schemes for Clocking/Timing of QCA Circuits," *IEEE Transactions on Computer-Aided Design of Integrated Circuits and Systems,* vol. 27, pp. 34-44, 2008.